\documentclass[12pt,fleqn]{article}
\usepackage{epsfig}

\begin{document}

%{\it final - for sub. - 16.X.2007}
%%%arXiv for published papers removed - 2.XII.2007

%24.VII.2008 erratum (linar weighting of b in UrQMD - replaced)

\begin{center}
  {\LARGE \bf Influence of Impact Parameter Fluctuations}
\end{center}
\begin{center}
 {\LARGE \bf on Transverse Momentum Fluctuations}
\end{center}

\vspace{0.5cm}

\begin{center}
  {\Large \bf Katarzyna Grebieszkow} \\
        Warsaw University of Technology\\
        e-mail: \tt{kperl@if.pw.edu.pl}
\end{center}

\vspace{0.5cm}

\begin{abstract}

The preliminary NA49 results on the energy dependence of transverse 
momentum fluctuations over the whole Super Proton Synchrotron (SPS) 
energy range exhibit an unexpected effect. The $\Phi_{p_{T}}$ 
fluctuation measure - used by the NA49 experiment - manifests a 
different behavior for different charge combinations. Whereas the 
$\Phi_{p_{T}}$ is consistent with zero and independent of energy for 
negatively charged particles, it significantly increases for both all 
charged and positively charged particles at lower SPS energies. The 
string-hadronic ultrarelativistic quantum molecular dynamics model (UrQMD) 
is applied here to explain this effect. The UrQMD simulations 
show that the number of protons is strongly correlated 
with impact parameter and that the event-by-event impact parameter 
fluctuations are responsible for the event-by-event transverse momentum 
fluctuations of positively charged and all charged particles where 
protons are included. The observations presented in this article are important 
for all experiments measuring event-by-event fluctuations, especially 
for those using all charged particles in the analysis. The 
results can be also crucial for detectors with acceptances extending to the 
beam/target spectator domains. 

\end{abstract}

\newpage

\section{Introduction}

\indent
Event-by-event fluctuations of kinematic characteristics and particle 
yields are believed to be one of the important probes to study the 
dynamics of heavy-ion collisions. Transverse momentum and multiplicity 
fluctuations are expected to be modified when the system approaches 
the phase boundary between hadron gas and quark-gluon plasma (QGP). It 
has been also argued that significant transverse momentum and 
multiplicity fluctuations should appear for systems hadronizing near the 
second-order critical QCD end-point \cite{SRS}. The QCD phase diagram - 
($T, \mu_B$), where $T$ is the temperature and $\mu_B$ bariochemical 
potential - can be scanned both by varying the system size/centrality 
and energy and therefore a possible nonmonotonic evolution of 
event-by-event ($p_T$ and multiplicity) fluctuations with beam energy, 
system size, or centrality may be used as an indication of the phase 
transition and the QCD critical point \cite{SRS}.

\indent
The NA49 experiment at the CERN SPS studied both the system size 
dependence and the energy dependence of transverse momentum 
fluctuations. In the analysis of the NA49 data the $\Phi_{p_T}$ 
fluctuation measure - proposed in \cite{Gaz92} - was used to quantify 
transverse momentum fluctuations on event-by-event basis (the 
definition and some properties of the $\Phi_{p_T}$ measure are presented 
in the next section and in \cite{Gaz92, fluct_size}). The NA49 results 
exhibit a significant nonmonotonic structure of transverse momentum 
fluctuations with the system size at the top SPS energy 
\cite{fluct_size}. Qualitatively the same structure was found for all 
possible charge selections i.e. for all charged, negatively charged, 
and positively charged particles. 

\indent
However, the preliminary NA49 results on the energy dependence 
of transverse momentum fluctuations over the whole SPS energy range 
manifests an unexpected effect: the $\Phi_{p_{T}}$ measure shows a 
different behavior for different charge selections (see \cite{cpod_kg} 
and Fig. \ref{fipt_URQMD_onlyone}). The $\Phi_{p_{T}}$ is independent 
of energy and it is consistent with zero for negatively charged 
particles, but it significantly increases for lower SPS energies 
for both all charged and positively charged particles.

\indent
In this work it will be shown that the effect observed by the NA49 
experiment is connected with protons only, and it can be explained by 
event-by-event impact parameter fluctuations or -
more precisely - by a correlation between the number of protons
in the forward hemisphere and the number of protons that are closer to 
the production region. The final conclusions from this analysis are 
important for all experiments measuring event-by-event fluctuations, 
especially for those that are using all charged particles in the 
analysis. It is also important for detectors with acceptances 
extending to the beam/target spectator domain. Due to a specific 
choice of the rapidity region (midrapidity only) the Relativistic 
Heavy Ion Collider (RHIC) experiments appear to be rather insensitive to 
the effects of impact parameter fluctuations - even if the analysis is 
performed by use of all charged particles registered in the detectors.

\indent
The analysis presented in this article is based on preliminary NA49 
results \cite{cpod_kg} and on events generated within the UrQMD 
approach \cite{urqmd1, urqmd2, urqmd3}.

\section{$\Phi_{p_T}$ fluctuation measure}

Following the authors of \cite{Gaz92} one defines a single-particle 
variable $z_{p_{T}}=p_{T}-\overline{p_{T}}$  
with the overbar denoting averaging over a single-particle inclusive 
distribution. Further, one introduces the event variable $Z_{p_{T}}$, 
which is a multiparticle analog of $z_{p_{T}}$, defined as
\begin{equation}
Z_{p_{T}}=\sum_{i=1}^{N}(p_{Ti}-\overline{p_{T}}),
\end{equation}
where the summation runs over particles in a given event. Note, that 
$\langle Z_{p_{T}} \rangle = 0$, where $\langle ... \rangle$ represents 
averaging over events. Finally, the $\Phi_{p_{T}}$ measure is defined 
as

\begin{equation}
\label{Phi}
\Phi_{p_{T}}=\sqrt{\frac{\langle 
Z_{p_{T}}^{2} \rangle }{\langle N
\rangle }}-\sqrt{\overline{z_{p_{T}}^{2}}},
\end{equation}
where $N$ is the average multiplicity of the considered particles. 
There are two most important properties of the $\Phi_{p_{T}}$ measure. 
If the system consists of particles that are emitted 
independently from each other (no interparticle correlations) 
$\Phi_{p_{T}}$ equals zero. On the other hand, if A+A is an incoherent 
superposition of many independent N+N interactions (superposition 
model), then $\Phi_{p_{T}}$ is independent of centrality and it has the 
same value for A+A and N+N collisions. 

\indent
Several effects may lead to nonzero value of $\Phi_{p_{T}}$. Among 
them are those that occur on an event-by-event basis (event-by-event 
fluctuations of the inverse slope parameter, existence of different 
event classes, i.e., ``plasma'' and ``normal'' events), but there are also 
interparticle correlations due to Bose-Einstein statistics, Coulomb 
effects, resonance decays, flow, jet production, etc.

\section{NA49 data}
\label{s:data}

\indent
To study the energy dependence of transverse momentum fluctuations the NA49 
experiment used samples of central Pb+Pb collisions at 20$A$, 30$A$, 
40$A$, 80$A$ and 158$A$ GeV energy, corresponding to $\sqrt{s_{NN}}$ = 
6.27, 7.62, 8.73, 12.3, and 17.3 GeV, respectively. The fraction of 
the total inelastic cross section of nucleus+nucleus collisions 
($\sigma/\sigma_{tot}$) was set as equal to 7.2\%. In the analysis only 
tracks with $0.005 < p_T < 1.5$ GeV/c were used. For all five SPS energies the 
forward-rapidity region was selected as $1.1 < y^*_{\pi} < 2.6$, where 
$y^*_{\pi}$ is the particle rapidity calculated in the center-of-mass 
reference system (particles were not identified in this analysis and 
their rapidities were calculated assuming the pion mass for all particles).
Within the studied rapidity range the azimuthal angle acceptance of the 
NA49 detector - ($p_T, \phi$) - is not uniform and it is 
described by the analytical curves given in \cite{cpod_kg}. 
Additionally, it was demonstrated \cite{cpod_kg} that at lower SPS 
energies, the NA49 TPC acceptance extends to the projectile spectator 
domain and therefore the sample of particles can be contaminated by 
beam particles. For energies above 80$A$ GeV the beam region and the 
region of particles - accepted by the NA49 for $p_T$ fluctuations 
analysis - do not overlap.

\begin{figure}[h]
%\vspace{-0.5cm}
\begin{center}
\epsfig{file= 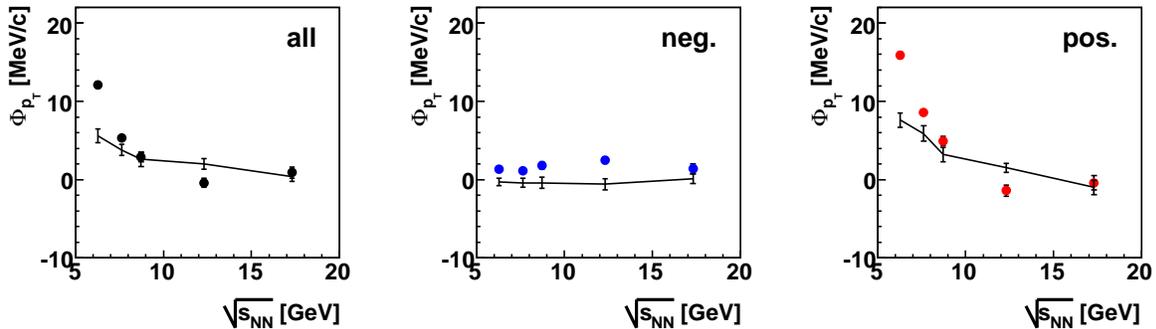,width=16cm}
\end{center}
\vspace{-0.5cm}
\caption {\small {(Color online) $\Phi_{p_{T}}$ as a function of energy
calculated from the UrQMD model (lines) with the acceptance
restrictions the same as for the preliminary NA49 data \cite{cpod_kg}.
UrQMD calculations are compared to preliminary NA49 data (points) taken
from \cite{cpod_kg}. The panels represent: all charged,
negatively charged, and positively charged particles, respectively. }}
\label{fipt_URQMD_onlyone}
\end{figure}

\indent
The preliminary NA49 results on the energy dependence of transverse 
momentum fluctuations are presented in Fig. \ref{fipt_URQMD_onlyone} 
(points). Separate panels correspond to all charged
particles, negatively charged and positively charged particles used in
the analysis. A significant decrease of transverse momentum fluctuations 
with increasing energy is observed for positively charged particles, 
whereas for the negatively charged ones $\Phi_{p_{T}}$ values are 
independent of energy.

\section{UrQMD predictions}

\indent
The $\Phi_{p_{T}}$ values measured by the NA49 experiment are compared 
to the predictions of the UrQMD model \cite{urqmd1, urqmd2, urqmd3}. 
The UrQMD generator is a microscopic transport model producing 
hadrons via formation, decay, and rescattering of resonances 
and strings. The UrQMD approach simulates multiple interactions of 
both target/beam nucleons and newly produced particles, excitation, 
and fragmentation of color strings and the formation and decay of 
hadronic resonances. In this analysis default parameters of the model have 
been used (meson-meson and meson-baryon scattering included). For each 
energy, from the sample of minimum bias Pb+Pb events, the most central 
7.2\% interactions have been selected, similarly as in the real NA49 
events. Such selection corresponds to the impact parameter range $0 < b 
< 4.35$ fm (in the generated histogram of impact parameter values the 
number of entries $N(b)$ was set as proportional to $const \cdot b$ 
[fm]).

\indent
The NA49 experiment used all charged particles, originating from the 
main vertex, to determine the $\Phi_{p_{T}}$ measure.
It means practically that only main vertex pions, protons, and kaons 
and their antiparticles were used in the analysis, because particles 
coming from the decays of $\Lambda$, $\phi$, $\Xi$ and $\Omega$ 
are believed to be rejected by a set of track selection criteria. 
Therefore the analysis of the UrQMD events has been carried out also by use of 
all charged pions, protons, and kaons and their antiparticles. In the 
analysis of the UrQMD events the same kinematic and acceptance 
restrictions have been applied as in the case of the NA49 data. In 
principle, the selected $p_T$ - azimuthal angle acceptance ($p_T, 
\phi$) is common for all five SPS energies and it corresponds to the 
restrictions used in the NA49 analysis (for details see \cite{cpod_kg}).

\indent
Fig. \ref{fipt_URQMD_onlyone} presents the comparison of the
$\Phi_{p_{T}}$ versus energy for the preliminary NA49 data (points) and 
for the UrQMD model (lines). Both the preliminary NA49 data and the 
UrQMD model show qualitatively the same behavior. The UrQMD model 
confirms a significant decrease of transverse momentum fluctuations 
with increasing energy for positively charged particles. On the other 
hand, there is no energy dependence of $\Phi_{p_{T}}$ for negatively 
charged particles. In the previous article of 
the NA49 \cite{fluct_size} the system size dependence of transverse 
momentum fluctuations was studied and a nonmonotonic structure of 
$\Phi_{p_{T}}$ appeared, but the shape of this dependence was very 
similar for all three charge selections. Therefore, the fact that the 
energy dependence of $\Phi_{p_{T}}$ is qualitatively different for 
various charge selections seems to be quite surprising. As the 
$\Phi_{p_{T}}$ value for all charged particles can be treated as a 
nontrivial combination of $\Phi_{p_{T}}$ for negatively and positively 
charged particles (additional sources of fluctuations, however, are not
excluded), a main emphasis should be placed on understanding the
qualitative difference between positively and negatively charged
particles.

\indent
It should be also stressed that, - although the same
kinematic and acceptance restrictions are used for the preliminary 
NA49 data and for the UrQMD events, - the $\Phi_{p_{T}}$ values should 
not be directly compared in these two cases. One of the reasons is a 
fact that the UrQMD model does not include effects of short-range 
correlations (Bose-Einstein and Coulomb). Moreover, it has been already 
shown by the CERES experiment that the measured values of mean $p_T$ 
fluctuations, calculated by use of the UrQMD model with default 
parameters, can be underestimated \cite{CERES} and that switching off
secondary scattering (in particular meson-baryon rescattering) results in
the values of $p_T$ fluctuations similar to those observed in the data.
It has been also checked that this underestimation (when rescattering
processes are included in the UrQMD model) can be even stronger for
lower energies. The above effect, observed by the CERES experiment,
might help in a qualitative explanation of the underestimation of the 
UrQMD $\Phi_{p_{T}}$ values in Fig. \ref{fipt_URQMD_onlyone}.

\section{Results for identified particles}

\begin{figure}[h]
%\vspace{-0.5cm}
\begin{center}
\epsfig{file= 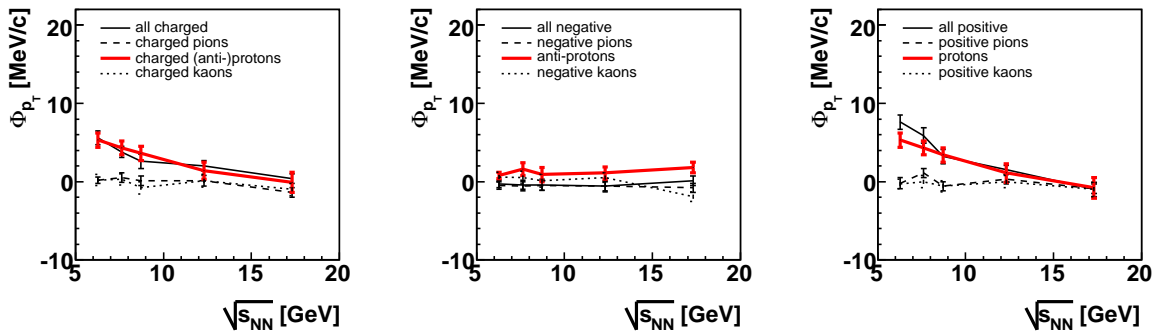,width=16cm}
\end{center}
\vspace{-0.5cm}
\caption {\small {(Color online) $\Phi_{p_{T}}$ versus energy calculated
from the UrQMD model, separately for different types of produced 
particles. The kinematic cuts and acceptance restrictions for the UrQMD 
model are the same as for the preliminary NA49 data. Pion mass has been 
assumed for all produced particles.}}
\label{fipt_URQMD_pi_p_k}
\end{figure}

\indent
The similar structure of the energy dependence of the
$\Phi_{p_{T}}$ measure for the NA49 data and for the UrQMD model
encourages us to search for an explanation of the 
origin of the increased 
$\Phi_{p_{T}}$ values for positively charged particles at lower SPS 
energies by analyzing effects incorporated in the model simulations. 
Although the track-by-track identification is not 
always possible in the data it is instructive to check - by use of 
the UrQMD model - what is the particle content in 
the studied forward-rapidity region. The NA49 sample of negatively 
charged particles is composed mainly of 
negative pions (94.4\% of all negatives for 20$A$ GeV and 89.4\% of all 
negatives for 158$A$ GeV) and negatively charged kaons (5.6\% for 20$A$ 
GeV and 9.2\% for 158$A$ GeV), whereas the number of 
antiprotons can be treated as negligible (below 0.1\% for 20$A$ GeV and 
1.6\% for 158$A$ GeV). The sample of positively charged particles is 
less homogeneous, as it contains positive pions (42.8\% for 20$A$ 
GeV and up to 69\% for 158$A$ GeV), positive kaons (from 8.5\% for 
20$A$ GeV to 10.4\% for 158$A$ GeV) and protons (48.7\% for 20$A$ GeV 
and 20.6\% for 
158$A$ GeV). Within the NA49 kinematic and acceptance region the total 
multiplicity of the final state protons remains nearly constant for all 
five energies (11.1 - 13.2), in contrary to the multiplicities of newly 
produced (generated) particles such as $\pi^+$ or $K^+$. Therefore, the 
fraction of protons in a sample of positively charged particles is 
significantly higher for lower energies and thus one can expect that the 
increased $\Phi_{p_{T}}$ values for lower energies might be indeed 
due to protons. The qualitative verification of this 
hypothesis is presented in Fig. \ref{fipt_URQMD_pi_p_k} where the 
$\Phi_{p_{T}}$ values are presented separately for different types of 
final state particles. The right panel of Fig. \ref{fipt_URQMD_pi_p_k} 
shows that the $\Phi_{p_{T}}$ values for protons only (solid, thick 
curve) are  very close to those for all positively charged particles 
(solid, thin curve) thus confirming 
that protons indeed are responsible for the increased $\Phi_{p_{T}}$ 
values at lower energies. The $\Phi_{p_{T}}$ measure obtained for newly 
produced particles such as pions, kaons (positively and negatively 
charged) or antiprotons is consistent with zero for all energies 
(in agreement with the hypothesis of weak interparticle correlations).

\begin{figure}[h]
%\vspace{-0.5cm}
\begin{center}
\epsfig{file= 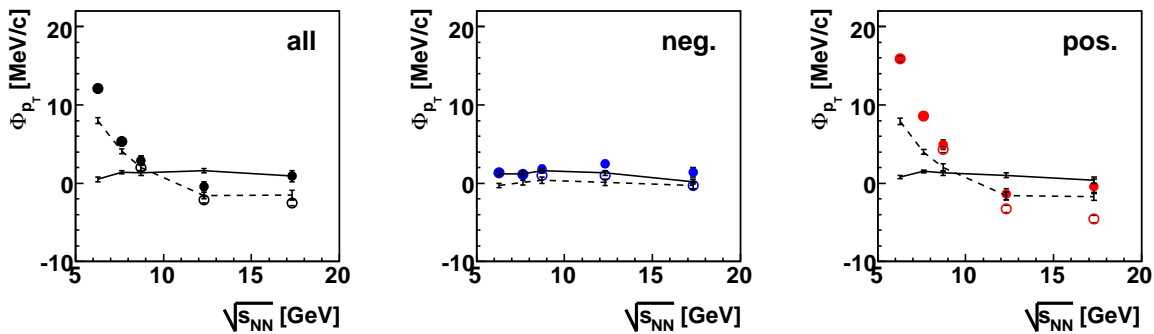,width=16cm}
\end{center}
\vspace{-0.5cm}
\caption {\small {(Color online) The preliminary NA49 data on
$\Phi_{p_{T}}$ as a function of energy calculated for unidentified
particles without two-track resolution corrections (open points) and
with two-track resolution corrections (full points)
compared to identified pions (solid curves) and (anti-)protons (dashed
curves). Data for identified particles do not include two-track
resolution corrections. For data with identification additional cuts on
$dE/dx$ values and total momentum ($p \geq 3$ GeV/c) were applied.
The panels represent: all charged, negatively charged, positively 
charged particles, respectively. Figure taken from \cite{cpod_kg}.}}
\label{fipt_energy_identif_p_cut}
\end{figure}

\indent
Although the precise track-by-track identification in the NA49 
experiment is much less reliable than the statistical one, the 
similar analysis with identified particles was prepared also for the 
NA49 data. In order to identify particles the combined information on 
the ionization energy loss ($dE/dx$) and total momentum (Bethe-Bloch curves) 
was used. The energy dependence of $\Phi_{p_{T}}$ measure was calculated
separately for charged pions and protons. Fig. 
\ref{fipt_energy_identif_p_cut} shows that the obtained NA49 results are 
in qualitative agreement with those from the UrQMD model, namely
$\Phi_{p_{T}}$ for pions (all charged and negatively and positively
charged separately: solid curves) and for antiprotons 
(dashed curve in the middle panel) are consistent with zero,
whereas $\Phi_{p_{T}}$ versus energy for protons only (dashed curve in 
the right panel) follows a structure observed for protons in the UrQMD 
model (strong decrease with energy). The quantitative comparison of 
$\Phi_{p_{T}}$ values - for positively charged particles and for protons 
only - is rather difficult because Bethe-Bloch identification leads 
to random losses of particles. The latter was found \cite{fluct_size} 
to reduce strongly the magnitude of correlations. Also an additional cut 
on total momentum ($p \geq 3$ GeV/c), applied to eliminate the region of 
intersection of Bethe-Bloch curves, reduces the magnitude of $\Phi_{p_{T}}$ 
values.

\section{Effect of the limited acceptance}

\begin{figure}[h]
%\vspace{-0.5cm}
\begin{center}
\epsfig{file= 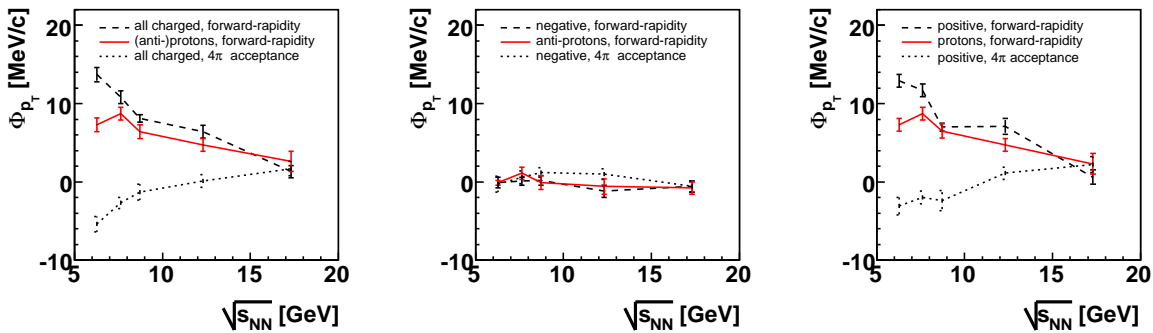,width=16cm}
\end{center}
\vspace{-0.5cm}
\caption {\small {(Color online) $\Phi_{p_{T}}$ versus energy calculated
from the UrQMD model without ($p_T - \phi$) acceptance restrictions 
(applied by the NA49). Pion mass has been assumed for all produced 
particles. The dashed and solid lines correspond to forward-rapidity 
region ($1.1<y^{*}_{\pi}<2.6$), and the dotted lines to `$4\pi$' acceptance. 
The panels represent: all charged, negatively charged and positively 
charged particles, respectively. Forward-rapidity (anti-)protons only 
are represented by solid lines.}}
\label{fipt_URQMD_4pi}
\end{figure}

\indent
It has been already stressed in many articles, that a {\it limited} 
acceptance can influence the observed values of fluctuation measures. 
(The NA49 experiment measured $p_T$ fluctuations with a specific choice of 
azimuthal angle - ($p_T, \phi$) curves defined in \cite{cpod_kg}.)  
Therefore, as the next step, it should be verified whether the energy 
dependence of 
$\Phi_{p_{T}}$ exhibits qualitatively the same structure when using 
complete azimuthal angle acceptance. The dashed lines in Fig. 
\ref{fipt_URQMD_4pi} represent the energy dependence of $\Phi_{p_{T}}$ 
for forward-rapidity region, when no ($p_T, \phi$) acceptance 
restrictions have been used. It is evident that the qualitative
structure of the energy dependence of $\Phi_{p_{T}}$ is very similar
also when using complete azimuthal angle acceptance. The $\Phi_{p_{T}}$
values have been also calculated using forward-rapidity (anti-)protons
only - also without ($p_T, \phi$) acceptance restrictions (solid
lines in Fig. \ref{fipt_URQMD_4pi}). Again, (anti-)protons follow the
behavior observed in Fig. \ref{fipt_URQMD_pi_p_k}, thus indicating that
the limited ($p_T, \phi$) acceptance indeed does not influence the 
observed effects significantly. In the studied forward-rapidity region 
(with $0.005 < p_T < 1.5$ GeV/c cut) the total multiplicity of final 
state protons is similar for all five energies and varies between 42.4 and 
47.6. Additionally, the $\Phi_{p_{T}}$ values have been evaluated for 
the so-called `$4\pi$' acceptance which means that only transverse 
momentum 
cut ($0.005 < p_T < 1.5$ GeV/c) has been used in the analysis (dotted 
lines in  Fig. \ref{fipt_URQMD_4pi}). However, in this ``complete'' 
kinematic acceptance the observed effect of interest seems to disappear. 
Instead, very small but negative $\Phi_{p_{T}}$ values are obtained 
for lower SPS energies. 

\begin{figure}[h]
%\vspace{-0.5cm}
\begin{center}
\epsfig{file= 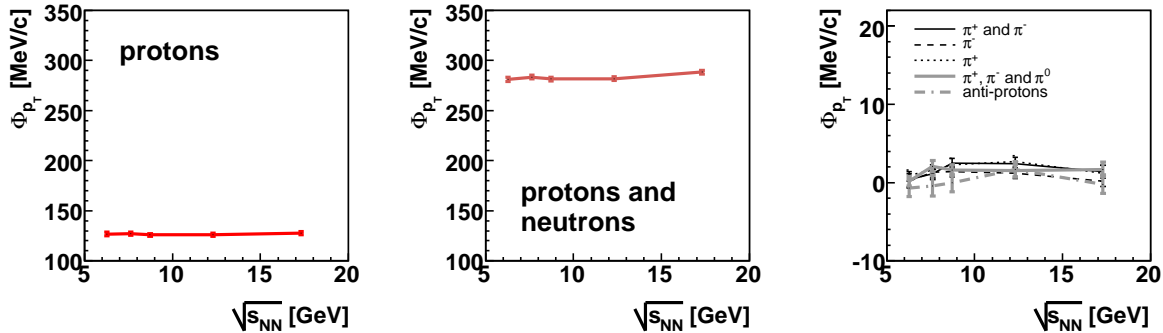,width=16cm}
\end{center}
\vspace{-0.5cm}
\caption {\small {(Color online) $\Phi_{p_{T}}$ versus energy 
calculated from the UrQMD model for `$4\pi$' acceptance ($0.005 <p_T < 
1.5$ GeV/c) for protons only (left), protons and neutrons (middle), and 
different selections of newly produced particles (right). }}
\label{fipt_URQMD_4pi_protons_neutrons}
\end{figure}

\indent
So far, one has learned that the effect of increased transverse momentum 
fluctuations arises probably from protons only. Now, one can try to 
establish whether those are newly produced protons, spectators or 
participants. It has been shown in \cite{cpod_kg} that for 
lower SPS energies the NA49 acceptance for $p_T$ fluctuations analysis 
extends to the beam spectator domain. Therefore a sample of particles 
used in the analysis can be 
contaminated by beam particles. Thus one can speculate that the increase 
of $\Phi_{p_{T}}$ values may be somehow attributed to the presence of
beam protons. Then, the measured correlation should be even
stronger for the detector with complete (`$4\pi$') acceptance because
some of the beam particles - spectators or elastically rescattered 
protons - might hit into the midrapidity region or 
even into the backward hemisphere. Moreover, also participating neutrons 
(as neutral particles are not registered in the detector) should increase 
the total $\Phi_{p_{T}}$ value. This hypothesis can be easily tested by 
use of the UrQMD events. The results of these calculations are 
presented in Fig. \ref{fipt_URQMD_4pi_protons_neutrons}, where $\Phi_{p_{T}}$
is calculated for `$4\pi$' acceptance ($0.005 <p_T < 1.5$ GeV/c) for
protons only (left) and for protons and neutrons (middle). The total
multiplicity of protons is very similar for all five energies and varies
between 167.1 and 169.9, whereas the total multiplicity of a 
mixture of protons with neutrons varies between 358.9 and 366.1. 
%(it is worth of pointing out that
%the estimated number of wounded nucleons for 7.2\% most central Pb+Pb
%interactions equals approximately 349). 
The very strong fluctuations observed for protons only ($\Phi_{p_{T}}$ 
at the level of 130 MeV/c) are exactly two times enlarged when neutrons 
are included. However, the $\Phi_{p_{T}}$ values for several 
selectios of newly produced particles such as pions or antiprotons remain
very small even for `$4\pi$' acceptance (Fig.
\ref{fipt_URQMD_4pi_protons_neutrons} - right), although the total
multiplicity of all pions is much higher than the multiplicity of 
protons and neutrons together and strongly depends on the energy. This 
result seems to indicate that the dominant effect originate not from 
the newly produced particles but from beam/target nucleons.

\section{Correlation between the spectator and production regions}

\begin{figure}[h]
%\vspace{-0.5cm}
\begin{center}
\epsfig{file= 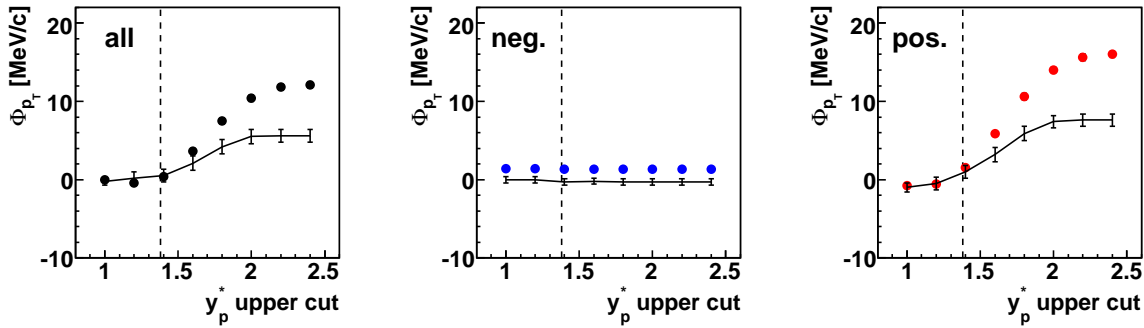,width=16cm}
\end{center}
\vspace{-0.5cm}
\caption {\small {(Color online) The NA49 and UrQMD results on
$\Phi_{p_{T}}$ as a function of an upper $y^{*}_{p}$ cut, obtained for
20$A$ GeV interactions. Points represent the NA49 data with
kinematic and acceptance restrictions as described in \cite{cpod_kg} and
Sec. \ref{s:data} (without $y_{p}^{*}$ cut). Solid lines correspond to
the UrQMD model with the acceptance restrictions the same as for the NA49
data. The NA49 data points are {\it not} corrected for limited two-track
resolution. The panels represent: all charged, negatively charged,
positively charged particles, respectively. For 20$A$ GeV interactions
the beam rapidity expressed in the center-of-mass reference system
$y^{*}_{beam}$=1.88. Note: the values and their errors are correlated.
The dashed lines indicate the $y_{p}^{*}$ cut finally used in the
analysis of the NA49 20$A$ GeV data. Figure taken from \cite{cpod_kg}.}}
\label{fipt_yproton_cut}
\end{figure}

\indent
Within the NA49 acceptance (forward rapidity) the possible 
contamination from beam particles is the highest for 20$A$ GeV 
interactions and therefore a test was carried out to check
how the $\Phi_{p_{T}}$ measure behaves when beam particles region is 
rejected. The NA49 results for 20$A$ GeV Pb+Pb collisions are 
presented in Fig. \ref{fipt_yproton_cut}, where the $\Phi_{p_{T}}$ 
measure was evaluated as a function of an upper $y^{*}_{p}$ cut 
($y^{*}_{p}$ is the particle rapidity calculated in the center-of-mass 
reference system assuming proton mass). 
In Fig. \ref{fipt_yproton_cut} the preliminary NA49 data are compared to 
predictions of the UrQMD model with the same acceptance restrictions. 
It is observed - both in the data and in the UrQMD model - that 
$\Phi_{p_{T}}$ decreases when the number of rejected ``beam particles'' 
increases.

\indent
It has been also observed that although $\Phi_{p_{T}}$ values measured
with exclusion of the beam spectator region are low, the $\Phi_{p_{T}}$
values obtained for ``spectator only'' region are also rather low,
whereas the whole magnitude of $p_T$ fluctuations can be reproduced only 
by using the whole rapidity interval. It suggests that the source of
fluctuations is not concentrated at very forward rapidities only (beam 
spectator domain) but there might be a strong correlation between 
particles from beam spectator domain and particles at lower rapidities 
(even in the production region). The above effect has been observed both 
in the NA49 data and in the UrQMD events.

\begin{figure}[h]
%\vspace{-0.8cm}
\begin{center}
\epsfig{file= 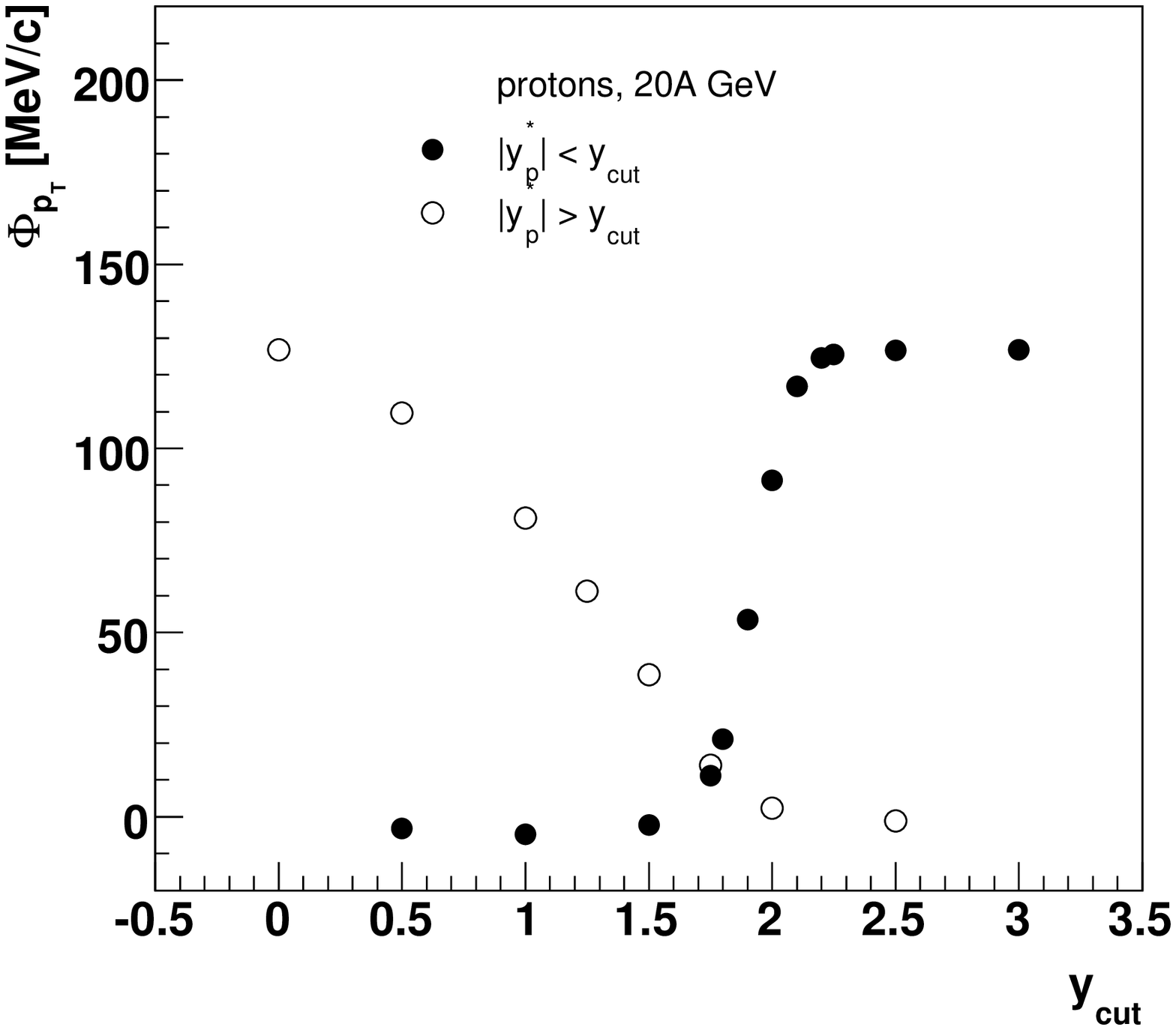,width=8cm}
\end{center}
\vspace{-1.4cm}
\caption {\small {$\Phi_{p_{T}}$ obtained for `$4\pi$' protons for the
UrQMD 7.2\% most central 20$A$ GeV interactions. The full points
represent the increasing acceptance around the midrapidity region. The
open points represent the increasing "gap" around midrapidity and
selection of particles closer to beam/target spectator regions. For
20$A$ GeV interactions the beam rapidity expressed in the center-of-mass
reference system $y^{*}_{beam}$=1.88. Note: the values and their errors
are correlated.}}
\label{back_forward}
\end{figure}

\indent
Figure \ref{back_forward} presents $\Phi_{p_{T}}$ obtained for `$4\pi$' 
protons for the UrQMD 20$A$ GeV interactions. The full points represent 
the increasing acceptance around the midrapidity region. The open points 
correspond to the increasing ``gap'' around midrapidity and thus 
selection of particles closer to beam/target spectator regions. Indeed, 
it can be observed that the maximal magnitude of $\Phi_{p_{T}}$ (about 
130 MeV/c) is obtained for a complete rapidity acceptance only, whereas 
$\Phi_{p_{T}}$ values either in the spectator domain or in the 
midrapidity region are low.

\section{The origin of correlation: event-by-event impact 
parameter fluctuations}

\indent
The natural explanation of the effect described above consists in 
event-by-event impact parameter fluctuations or - more precisely - in 
correlation between the number of protons (nucleons) in the forward 
hemisphere and the number of protons (nucleons) in the production 
region. For more central events the number of forward-rapidity protons 
is smaller, and consequently, the number of protons in the production 
region is higher. The situation is opposite for less central 
collisions. The existence of those different event classes results in 
the increased $\Phi_{p_{T}}$ values for positively charged particles.

\begin{figure}[h]
%\vspace{-0.8cm}
\begin{center}
\epsfig{file= 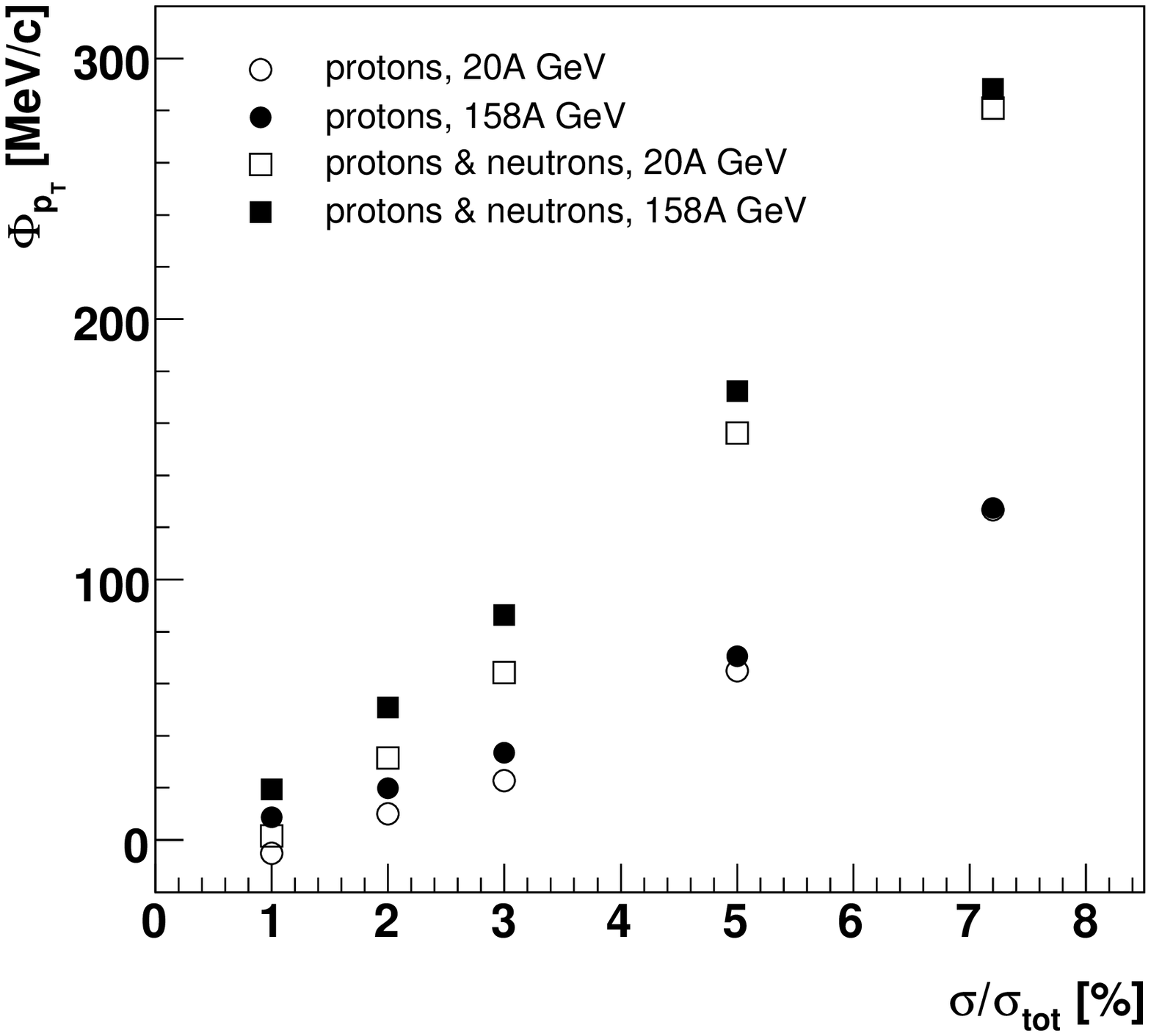,width=8cm}
\end{center}
\vspace{-1.4cm}
\caption {\small {$\Phi_{p_{T}}$ as a function of the fraction of the
total inelastic cross section of nucleus+nucleus collisions
($\sigma/\sigma_{tot}$). The results are obtained for `$4\pi$' protons
or for a mixture of protons and neutrons for 20$A$ GeV and 158$A$ GeV
UrQMD events. Note: the values and their errors are correlated.}}
\label{pn_percent}
\end{figure}

\indent
A test of the hypothesis of the correlation between 
protons in different rapidity windows can be found in Fig. 
\ref{pn_percent}. Two kinds of the UrQMD particle samples: protons only 
and a mixture of protons and neutrons were used to obtain the 
$\Phi_{p_{T}}$ dependence on the fraction of the total inelastic cross
section of nucleus+nucleus collisions ($\sigma/\sigma_{tot}$). The UrQMD 
calculations correspond to `4$\pi$' acceptance in both 20$A$ and 158$A$ 
GeV Pb+Pb interactions. The UrQMD results show a significant reduction 
of $\Phi_{p_{T}}$ values when the centrality is restricted.
The original $\Phi_{p_{T}}$ value for protons only at the level of 130 
MeV/c (for 7.2\% most central interactions at 20$A$ GeV) is reduced 
to the value consistent with zero when the centrality is restricted to 
1\% most central Pb+Pb interactions. 

\begin{figure}[h]
%\vspace{-0.5cm}
\begin{center}
\epsfig{file= 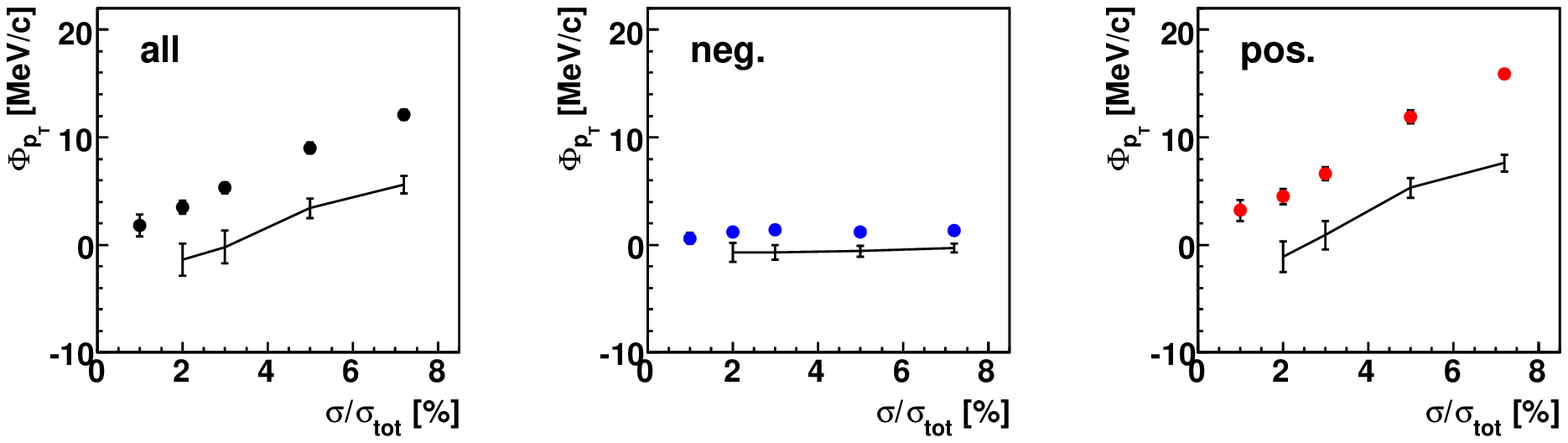,width=16cm}
\end{center}
\vspace{-0.5cm}
\caption {\small {(Color online) The NA49 and UrQMD results on
$\Phi_{p_{T}}$ as a function of the fraction of the total inelastic
cross section of nucleus+nucleus collisions ($\sigma/\sigma_{tot}$) for
20$A$ GeV Pb+Pb collisions. Points represent NA49 data with kinematic
and acceptance cuts as described in Sec. \ref{s:data} (without
$y_{p}^{*}$ cut). Solid lines correspond to
the UrQMD model with the acceptance restrictions the same as for the
NA49 data. The NA49 data points are {\it not} corrected for limited two-track
resolution. The panels represent: all charged, negatively charged, and
positively charged particles, respectively. Note: the values and their
errors are correlated. Figure taken from \cite{cpod_kg}.}}
\label{fipt_percent_inel}
\end{figure}

\indent
The similar effect was also observed within the NA49 acceptance region. 
Figure \ref{fipt_percent_inel} presents the $\Phi_{p_{T}}$ values - for 
20$A$ GeV interactions - plotted as a function of the fraction of the total 
inelastic cross section ($\sigma/\sigma_{tot}$). The preliminary 
NA49 data were compared to the predictions of the UrQMD model with the 
same kinematic restrictions. Both the NA49 data (points) and the UrQMD 
events (lines) show a significant reduction of $\Phi_{p_{T}}$ values 
when the centrality is restricted to more central events.

\indent
The effect observed in Fig. \ref{pn_percent} encourages us to 
study minimum bias interactions where impact parameter fluctuations from 
event to event should be maximal. The $\Phi_{p_{T}}$ values, obtained for
minimum bias 20$A$ GeV UrQMD events (30k events), are demonstrated in 
Table \ref{minbias}. Results are shown for several particle selections 
at `4$\pi$' acceptance. Transverse momentum fluctuations for newly 
produced particles (pions) are consistent with zero, whereas for protons 
and for a sample of protons and neutrons - as it could be expected - 
exhibit extremely high 
$\Phi_{p_{T}}$ values. Relatively high $\Phi_{p_{T}}$ value can be also 
obtained when protons are combined with positively charged kaons and 
pions, whereas in the case of central data (Fig. 
\ref{fipt_URQMD_4pi}) at `4$\pi$' acceptance, the presence of 
pions and kaons in a sample turned out to be sufficient to
wash out the correlation originating from protons.

\begin{table}[h]
\begin{center}
\begin{small}
\begin{tabular}{|c|c|c|c|}
\hline
& $\Phi_{p_{T}}$ [MeV/c] & $\langle N \rangle$ & $\sigma _N$ \cr
\hline
\hline
$p$ & 1506 $\pm$ 5.0 & 167.1 & 6.2 \cr
\hline
$p$ and $n$ & 2230.9 $\pm$ 7.1 & 402.5 & 16.3 \cr
\hline
$\pi^{+}$ and $\pi^{-}$ & -0.3 $\pm$ 1.4 & 157.5 & 171.8 \cr
\hline
$\pi^{-}$ & 2.2 $\pm$ 1.5 & 83.5 & 90.7 \cr
\hline
$\pi^{+}$ & -1.2 $\pm$ 1.4 & 74.1 & 81.3 \cr
\hline
$\pi^{+}, \pi^{-}, p, \bar{p}, K^{+}, K^{-}$ & 1106.7 $\pm$ 4.7 & 335.8
& 187.1 \cr
\hline
$\pi^{-}, \bar{p}, K^{-}$ & 2.7 $\pm$ 1.5 & 86.1 & 93.8 \cr
\hline
$\pi^{+}, p, K^{+}$ & 1290.6 $\pm$ 4.1 & 249.6  & 93.3 \cr
\hline
\end{tabular}
\end{small}
\end{center}
\vspace{-0.5cm}
\caption {\small {$\Phi_{p_{T}}$ values, mean multiplicities, and their
dispersions ($\sigma_N=\sqrt{\langle N^2 \rangle -\langle N \rangle
^2}$) obtained for minimum bias 20$A$ GeV UrQMD events (30k events).
Results are shown for several particle combinations at `4$\pi$'
acceptance.}}
\label{minbias}
\end{table}

\indent
Figures \ref{back_forward} and \ref{pn_percent} together with the 
corresponding figures within the limited NA49 acceptance (Fig. 
\ref{fipt_yproton_cut} and Fig. \ref{fipt_percent_inel}) seem to
confirm the hypothesis that the increased $\Phi_{p_{T}}$ values can be
indeed connected with event-by-event fluctuations of the number of
protons in the forward hemisphere and the number of protons that 
are closer to the production region. There are at least two possible 
ways to eliminate this trivial source of correlations. The first - and 
the most natural one - consists in centrality restrictions, however it 
can significantly reduce the event statistics. The second one relies on 
applying an additional rapidity cut that would reject the beam spectator 
domain (see the vertical lines in Fig. \ref{fipt_yproton_cut}).

\begin{figure}
%\vspace{-0.5cm}
\begin{center}
\epsfig{file= 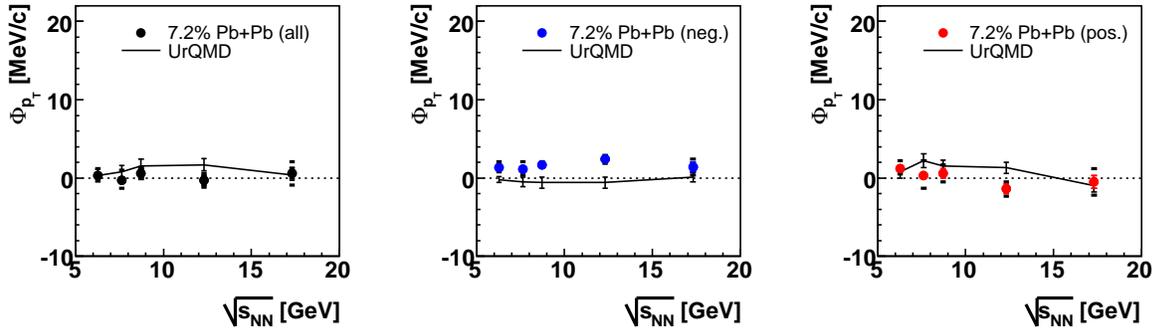,width=16cm}
\end{center}
\vspace{-0.5cm}
\caption {\small {(Color online) The NA49 and UrQMD results on
$\Phi_{p_{T}}$ as a function of energy for the 7.2\% most central Pb+Pb 
interactions for all charged particles (left) and for negatively charged 
(middle) and positively charged ones (right). The $\Phi_{p_{T}}$ values 
are obtained with additional cut: only particles obeying $y^*_p < 
y^{*}_{beam} - 0.5$. NA49 data points are shown with statistical and 
systematic errors. The NA49 results are compared to the UrQMD 
predictions (solid lines) with the same acceptance restrictions. The 
NA49 data points taken from \cite{cpod_kg}.}}
\label{fipt_energy1_RAP_CUT_3panels}
\end{figure}

\indent
The NA49 experiment chose the second (rejection) method and an 
additional $y_{p}^{*}$ cut was applied for each energy, namely 
for each produced particle its $y_{p}^{*}$ was required to be lower than 
$y^{*}_{beam}-0.5$, where $y^{*}_{beam}$ is the beam rapidity 
in the center-of-mass reference system. The final NA49 results on 
the $\Phi_{p_{T}}$ fluctuation measure, as a function of energy, are shown
in Fig. \ref{fipt_energy1_RAP_CUT_3panels}. Three panels represent all
charged, negatively charged, and positively charged particles,
respectively. The points correspond to the data (with statistical and
systematic errors) and the lines to the predictions of the UrQMD model. 
When additional cut on $y_{p}^{*}$ is applied, no significant energy 
dependence of $\Phi_{p_{T}}$ measure can be observed for all three charge
selections and $\Phi_{p_{T}}$ values are consistent with the hypothesis 
of independent particle production (that is $\Phi_{p_{T}} \approx 0$).

\section{Final-state protons in the UrQMD model}

\indent
The above analysis confirms that the preliminary NA49 results, i.e., the 
increased $\Phi_{p_{T}}$ values at lower SPS energies were indeed 
due to protons only. The basic origin of this effect is the 
correlation between the number of protons (generally nucleons) in the 
forward hemisphere and the number of protons closer to the production 
region, caused by event-by-event impact parameter fluctuations. At the 
end of this analysis it is worthwhile to find out what is the origin of 
the final-state protons in the UrQMD model. In principle, it should be 
clarified what kind of protons are rejected by an additional cut on 
$y_{p}^{*}$, applied by the NA49 experiment.

\begin{figure}[h]
%\vspace{-0.6cm}
\begin{center}
\epsfig{file= 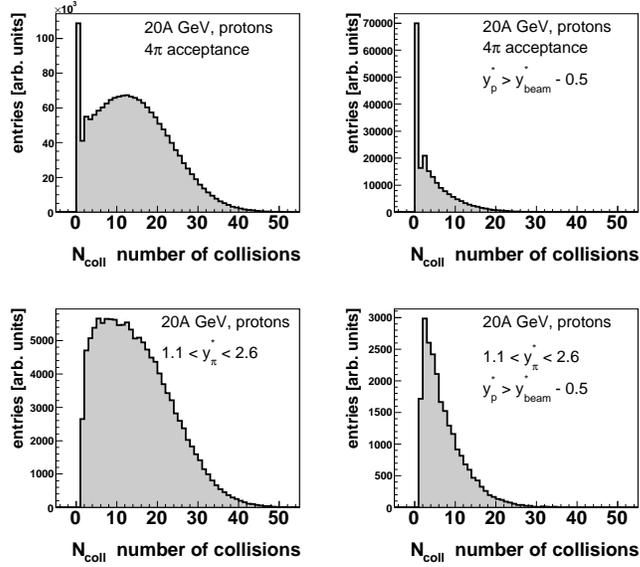,width=9cm}
\end{center}
\vspace{-0.8cm}
\caption {\small {The distributions of the number of collisions of 
final-state protons generated by the UrQMD model for the 7.2\% most central 
Pb+Pb interactions at 20$A$ GeV. Different panels correspond to 
different acceptance regions.}}
\label{colln}
\end{figure}

\indent
Figure \ref{colln} shows the distributions of the number of collisions 
of final state protons, generated by the UrQMD model for the 7.2\% most 
central Pb+Pb interactions at 20$A$ GeV. Four panels correspond to 
different acceptance regions. The peaks at zero (upper panels) represent 
real spectators, i.e., protons that do not experience any collision. 
Their fraction is obviously higher when one restricts 
rapidity acceptance to the regions closer to the rapidity of the beam 
(Fig. \ref{colln} upper, right). The mean number of collisions is 
generally smaller for the acceptance characteristic of the beam spectators 
(right panels). Fig. \ref{colln} (lower, left) shows protons in the 
acceptance selected by the NA49 experiment, whereas Fig. \ref{colln} 
(lower, right) - protons that were finally excluded from the NA49 
acceptance by an additional cut on $y_{p}^{*}$. It is quite surprising  
that in the two lower panels there are no spectators ($N_{coll}=0$) at all. 
It would suggest that the effect of increased $\Phi_{p_{T}}$ 
values was connected with participants only and that spectators did not 
play any important role. This is, however, true only inside the 
NA49 acceptance and it will be shown below (Table \ref{tabhistory}) that 
when the `4$\pi$' acceptance is considered, both real spectators 
and participants, i.e., elastically scattered protons (precisely, 
the correlation between multiplicities of those two) are responsible 
for the observed $\Phi_{p_{T}}$ values.  

\begin{figure}[h]
%\vspace{-0.6cm}
\begin{center}
\epsfig{file= 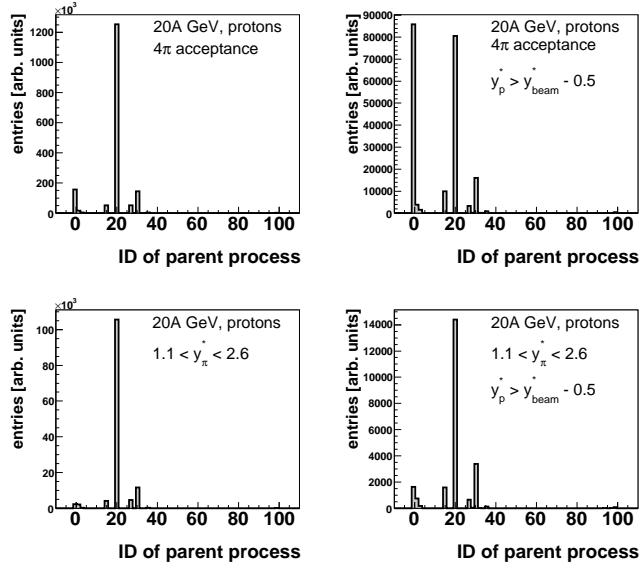,width=9cm}
\end{center}
\vspace{-0.8cm}
\caption {\small {The distributions of ID of parent process (last process
the particle was involved in) obtained for final-state protons generated 
by the UrQMD model for the 7.2\% most central Pb+Pb interactions at 
20$A$ GeV. Different panels correspond to different acceptance regions.}}
\label{history}
\end{figure}

\indent
During the collision process the projectile and target nucleons 
participate both in elastic and in inelastic interactions. In 
principle, $\Delta$ resonances can be produced as an effect of inelastic 
collisions. Afterward, $\Delta$s decay into pions and nucleons. 
Technically - within the UrQMD model - for each final state particle one 
can attribute an integer number called the "ID of parent process", 
which describes the last process the particle was involved in. Therefore, 
it is easy to disentangle spectator protons from those 
originating from elastic or inelastic processes.  
Figure \ref{history} presents the distributions of ID of parent process 
(last process the particle was involved in) obtained for the final-state 
protons generated in the UrQMD model for the 7.2\% most central Pb+Pb 
interactions at 20$A$ GeV. Different panels correspond to 
different acceptance regions - similarly to Fig. \ref{colln}. Below, 
one can find a list of IDs that significantly contribute to the histogram. 

\indent
The ID of parent process equal zero denotes real spectators 
($N_{coll}=0$) but also elastically scattered protons (scattering with 
all barions and mesons is possible). For protons with ID of parent 
process equal zero the number of collisions ($N_{coll}$) has a high 
peak at zero (real spectators) and a tail up to about 10-15, whereas 
the mean value of $N_{coll}$ is about 0.5. 
All protons originating from decays (for example, from resonances) has ID 
of parent process equals 20. As we are considering central 
collisions (7.2\%) the majority of protons come from decays with 
exception of very forward-rapidity regions where spectators begin to 
dominate (Fig. \ref{history} upper, right).  
The ID of parent process equals 1 means protons generated in the reaction 
$NN \rightarrow N \Delta $ \footnote{$N \equiv
N_{938}$ (nucleon), $N^{*} \equiv N_{1440}, N_{1520}, N_{1535}$, ...
(nucleon resonances), $\Delta \equiv \Delta_{1232}, \Delta^{*} \equiv
\Delta_{1600}, \Delta_{1620}, \Delta_{1700}, ... $}, ID equals 14 means 
protons produced in inelastic interactions (without string excitation) 
such as: $N^{*} N, \Delta^{*} N, \Delta N^{*}, 
\Delta^{*} N^{*}$, or $N^{*} N^{*}$. The parent history ID labeled as 
30 denotes process $\Delta N \rightarrow NN$. And, finally, IDs 
equal 15 and 27 correspond to processes through 2 strings and process 
through 1 string, respectively.

\indent
Table \ref{tabhistory} shows the $\Phi_{p_{T}}$ values for `4$\pi$' protons   
obtained for UrQMD 7.2\% most central Pb+Pb interactions at 20$A$ GeV, 
where several choices of number of collisions ($N_{coll}$) or ID of 
parent process have been applied. The $\Phi_{p_{T}}$ value obtained 
for spectators ($N_{coll}=0$) is low but it is low also when the 
spectators are rejected. The original magnitude of $\Phi_{p_{T}}$ 
(about 130 MeV/c) can be reproduced only when both spectators and ``not 
spectators'' are used in the analysis. It simply indicates that there 
should be some kind of correlation between spectators and ``not 
spectators''. The same observation can be achieved for ID of parent process 
equal and not equal zero. The opposite situation takes place when 
protons originating from decays are considered. The $\Phi_{p_{T}}$ value 
evaluated for protons coming from decays is rather low: however, if 
such protons are rejected from the analysis, - the $\Phi_{p_{T}}$ does 
not decrease but even increases! It means that protons from decays 
(i.e. newly produced particles) do not generate the dominant 
correlation in the analysis. It is in agreement with intuitive 
predictions, because also 
newly produced particles such as pions, kaons, or antiprotons lead to 
$\Phi_{p_{T}}$ value consistent with zero. Finally, it has been observed 
that not only protons from decays but also protons created in a list of 
inelastic processes (labeled as 1, 14, 15, 27, 30) result in a 
$\Phi_{p_{T}}$ values consistent with zero. Therefore it can be concluded 
that the whole effect of the increased transverse momentum fluctuations  
comes from beam/target particles only (both real spectators and 
elastically scattered beam/target protons). Those particles can hit different 
rapidity windows - depending on the centrality of a given event. 
Consequently, the sample composed by a mixture of events with 
different impact parameters generates increased event-by-event $p_T$
fluctuations. The impact parameter range in the 7.2\% most central Pb+Pb 
interactions is wide enough to produce significantly increased 
$\Phi_{p_{T}}$ values.   

\begin{table}[h]
\begin{center}
\begin{small}
\begin{tabular}{|c|c|c|c|}
\hline
& $\Phi_{p_{T}}$ [MeV/c] & $\langle N \rangle$ & $\sigma _N$ \cr

\hline
\hline

\begin{tabular}{c}
rejection of protons with $N_{coll} = 0$ \cr (spectators)
\end{tabular}
& 35.5 $\pm$ 1.4  & 158.2  & 10.3 \cr

\hline

\begin{tabular}{c}
only protons with $N_{coll} = 0$ \cr (spectators)
\end{tabular}
& 4.5 $\pm$ 0.2 & 10.8 & 6.0 \cr

\hline
\hline

\begin{tabular}{c}
rejection of protons with ID of parent process = 0 \cr (spectators and
elastically scattered protons)
\end{tabular}
& 16.3 $\pm$ 1.4 & 153.5 & 11.2 \cr

\hline

\begin{tabular}{c}
only protons with ID of parent process = 0  \cr (spectators and
elastically scattered protons)
\end{tabular}
& 5.8 $\pm$ 0.7 & 15.6 & 7.9 \cr

\hline
\hline

\begin{tabular}{c}
rejection of protons with ID of parent process = 20 \cr (protons from
decays)
\end{tabular}
& 156.7 $\pm$ 1.6 & 43.7 & 10.1 \cr

\hline

\begin{tabular}{c}
only protons with ID of parent process = 20 \cr (protons from decays)
\end{tabular}
& 12.6 $\pm$ 1.2 & 125.3 & 11.5 \cr

\hline
\hline

\begin{tabular}{c}
rejection of protons with ID of parent process = 1 \cr ($NN \rightarrow
N \Delta $ )
\end{tabular}
& 125.2 $\pm$ 1.5 & 167.4 & 9.2 \cr

\hline

\begin{tabular}{c}
only protons with ID of parent process = 1 \cr ($NN \rightarrow N
\Delta $)
\end{tabular}
& 1.9 $\pm$ 1.2 & 1.7 & 1.3 \cr

\hline
\hline

\begin{tabular}{c}
rejection of protons with ID of parent process = 14 \cr ($N^{*} N,
\Delta^{*} N, \Delta N^{*}, \Delta^{*} N^{*}$ or $N^{*} N^{*}$)
\end{tabular}
& 127.5 $\pm$ 1.4 & 166.3 & 9.2 \cr

\hline

\begin{tabular}{c}
only protons with ID of parent process = 14 \cr ($N^{*} N, \Delta^{*}
N, \Delta N^{*}, \Delta^{*} N^{*}$ or $N^{*} N^{*}$)
\end{tabular}
& 1.8 $\pm$ 1.6 & 2.7 & 1.7 \cr

\hline
\hline

\begin{tabular}{c}
rejection of protons with ID of parent process = 15 \cr (process trough
2 strings)
\end{tabular}
& 125.8 $\pm$ 1.4 & 166.5 & 9.2 \cr

\hline

\begin{tabular}{c}
only protons with ID of parent process = 15 \cr (process trough 2
strings)
\end{tabular}
& 1.3 $\pm$ 1.2 & 2.5 & 1.6 \cr

\hline
\hline

\begin{tabular}{c}
rejection of protons with ID of parent process = 27 \cr (process trough
1 string)
\end{tabular}
& 127.3 $\pm$ 1.4 & 163.9 & 9.2 \cr

\hline

\begin{tabular}{c}
only protons with ID of parent process = 27 \cr (process trough 1
string)
\end{tabular}
& 0.8 $\pm$ 1.0 & 5.1 & 2.3 \cr

\hline
\hline

\begin{tabular}{c}
rejection of protons with ID of parent process = 30 \cr ($\Delta N
\rightarrow NN$)
\end{tabular}
& 126.9 $\pm$ 1.4 & 155.0 & 9.1 \cr

\hline

\begin{tabular}{c}
only protons with ID of parent process = 30 \cr ($\Delta N \rightarrow
NN$)
\end{tabular}
& 2.7 $\pm$ 1.1 & 14.0 & 4.3 \cr

\hline
\hline

\end{tabular}
\end{small}
\end{center}
\vspace{-0.5cm}
\caption {\small {$\Phi_{p_{T}}$ values, mean multiplicities, and their
dispersions ($\sigma_N=\sqrt{\langle N^2 \rangle -\langle N \rangle
^2}$) obtained for UrQMD 7.2\% most central Pb+Pb interactions at
20$A$ GeV. Results are shown for protons at `4$\pi$' acceptance with
several choices of number of  collisions ($N_{coll}$) or ID of parent
process (see the text for details).}}
\label{tabhistory}
\end{table}

\clearpage

\section{Summary and Conclusions}

\indent
It has been presented how event-by-event impact parameter 
fluctuations influence transverse momentum fluctuations. In principle, 
it has been explained that the preliminary NA49 results - showing 
significantly increased $\Phi_{p_{T}}$ values at lower SPS energies -  
are connected with the trivial effects of impact parameter fluctuations.  
It suggests that a precise selection of very narrow centrality bins 
should be crucial not only for the analysis of multiplicity fluctuations 
\cite{mryb} but also for transverse momentum fluctuations and probably 
for other kinematic characteristics. 

\indent
It was predicted that either approaching the phase boundary of the QCD 
phase diagram or hadronization close to the critical point should lead to 
increased values of transverse momentum fluctuations. Therefore it is 
very important to distinguish between interesting physical 
effects and trivial ones as those connected with impact 
parameter fluctuations. Below, one can find several methods to eliminate 
the effects of event-by-event impact parameter fluctuations.

\begin{enumerate}
\item One can perform the analysis for all three charge 
selections (all charged, negatively charged, positively charged)
and check whether the same structure (not necessarily the magnitude!) 
of $\Phi_{p_{T}}$ versus energy or centrality (system size) appears 
for all possible charge selections (see the NA49 results in 
\cite{fluct_size}).

\item If the previous method is not possible it is safer to restrict the 
analysis to negatively charged particles only, instead of all charged or 
positively charged ones.

\item The experiments with reliable particle identification can try to 
reject protons from the sample and measure $\Phi_{p_{T}}$ for newly 
produced particles only (pions or a mixture of pions, kaons, etc.).

\item One can also probe the midrapidity region only (production 
domain) or at least one can check whether the selected acceptance does 
not extend to the beam/target spectator domain.

\item The best and the least controversial method relies on a drastic 
centrality restriction. However, due to insufficient statistics, this 
method cannot be applied by every experiment.
    
\end{enumerate}

It has been already mentioned that due to the specific choice of the 
rapidity region (midrapidity only) the RHIC experiments are essentially 
free of the effects of impact parameter fluctuations,  
even though in most cases the results are obtained by use of all 
charged particles registered in the detectors. The same remark concerns 
also the CERES experiment at CERN SPS that measured transverse momentum 
fluctuations in 40$A$, 80$A$, and 158$A$ GeV Pb+Au interactions by use of 
all charged particles - but produced in midrapidity region only 
\cite{CERES}. In contrary, the acceptance of the NA49 experiment - 
selected for the analysis of transverse momentum fluctuations - extends 
to the beam spectator domain. However, the NA49 experiment performs 
the analysis by use of all three charge selections and therefore it 
was possible to separate and finally eliminate the trivial effects 
connected with the impact parameter fluctuations.

\newpage

\noindent
{\bf Acknowledgments}

I am indebted to the authors of the UrQMD model for the permission to use 
their code in my analysis. I am very grateful to Marek Ga\'zdzicki for 
his interesting comments and suggestions concerning this analysis. I 
also appreciate the editorial remarks from Ewa Skrzypczak and Staszek 
Mr\'owczy\'nski. Finally I thank Jarek Grebieszkow for the software 
support.

%\newpage

\end{document}